\documentclass{iopart}
\usepackage[colorlinks=true,citecolor=blue]{hyperref}
\usepackage{graphicx}
\usepackage{iopams}

\newcommand{\GF}{G_{\mathrm{F}}} 
\newcommand{\diag}{\mathrm{diag}} 
\newcommand{\vac}{\mathrm{vac}} 
\newcommand{\matt}{\mathrm{matt}} 
\newcommand{\ext}{\mathrm{ext}} 
\newcommand{\tot}{\mathrm{tot}} 
\newcommand{\dm}{\Delta m^2} 
\newcommand{\bhp}{\hat{\bi{p}}}   
\newcommand{\ve}{\vec{e}}   
\newcommand{\vs}{\vec{s}}   
\newcommand{\vP}{\vec{P}}   
\newcommand{\vH}{\vec{H}}   
\newcommand{\vtH}{\vec{\tilde{H}}}   
\newcommand{\vtP}{\vec{\tilde{P}}}   
\newcommand{\bfx}{\bi{x}}   
\newcommand{\bfr}{\bi{r}}   
\newcommand{\bfK}{\bi{K}}   
\newcommand{\bfp}{\bi{p}}   
\newcommand{\sfH}{\mathsf{H}}   
\newcommand{\sfP}{\mathsf{P}}   
\newcommand{\sfU}{\mathsf{U}}   
\newcommand{\sfG}{\mathsf{G}}   
\newcommand{\stH}{\tilde{\mathsf{H}}}   
\newcommand{\stP}{\tilde{\mathsf{P}}}   
\newcommand{\tP}{\tilde{P}}   
\newcommand{\thetav}{\theta_\mathrm{v}}   
\newcommand{\bnab}{\boldsymbol{\nabla}} 
\newcommand{\Es}{E^\mathrm{s}} 

\begin{document}

\title{Symmetries in collective neutrino oscillations}

\author{H Duan$^1$, G M Fuller$^2$ and Y-Z Qian$^3$}
\address{$^1$ Institute for Nuclear Theory, %
University of Washington, %
Seattle, WA 98195, USA}
\address{$^2$ Department of Physics, %
University of California, San Diego, %
La Jolla, CA 92093, USA}
\address{$^3$ School of Physics and Astronomy, %
University of Minnesota, Minneapolis, MN 55455, USA}

\ead{\mailto{huaiyu.duan@mailaps.org},
\mailto{gfuller@ucsd.edu},
\mailto{qian@physics.umn.edu}}

\begin{abstract}
We discuss the relationship between a symmetry in the neutrino flavour
evolution equations and neutrino flavour oscillations in the collective
precession mode.
This collective
precession mode can give rise to spectral swaps (splits) when
conditions can be approximated as homogeneous and isotropic. 
Multi-angle numerical simulations of supernova neutrino flavour
transformation show that when this approximation breaks down,
non-collective neutrino oscillation modes decohere kinematically, but
the collective precession mode still is expected to stand out. 
We provide a criterion for significant flavour
transformation to occur if neutrinos participate in a collective precession
mode. This criterion can be used to understand the suppression of collective
neutrino oscillations in anisotropic environments in the presence of a high
matter density. This criterion is also useful in understanding the
breakdown of the collective precession mode when neutrino densities
are small.
\end{abstract}
 
\maketitle

\section{Introduction}

Because of neutrino-neutrino forward scattering or neutrino
self-interaction
\cite{Fuller:1987aa,Notzold:1988kx,Pantaleone:1992xh} neutrinos can
experience collective flavour transformation in environments 
such as the early Universe (e.g., \cite{Samuel:1993uw,Kostelecky:1994dt,Samuel:1995ri,Pastor:2001iu,Dolgov:2002ab,Abazajian:2002qx})
and supernovae
(e.g., \cite{Qian:1994wh,Pastor:2002we,Balantekin:2004ug,Fuller:2005ae})
where neutrino number densities can be very large.
This phenomenon is different
from the conventional Mikheyev-Smirnov-Wolfenstein (MSW) effect
\cite{Wolfenstein:1977ue,Mikheyev:1985aa} in that
the flavour evolution histories of neutrinos in collective oscillations
are coupled together and must be solved simultaneously. 
The possibility and consequences of collective neutrino oscillations in
supernovae were not well appreciated until the discovery that the ordinary
matter can be ``ignored'' in such phenomena \cite{Duan:2005cp} and the
first numerical demonstrations of ``stepwise spectral swapping'' 
(or ``spectral split'') \cite{Duan:2006jv,Duan:2006an} which is the
imprint left by the collective flavour transformation on neutrino
energy spectra. 

Significant progress has been made towards understanding collective
neutrino oscillations in supernovae (see, e.g., \cite{Duan:2009cd} for a
brief review and references therein). In particular, Raffelt and
Smirnov \cite{Raffelt:2007cb} demonstrated an adiabatic (precession)
solution for the homogeneous, isotropic neutrino gas.
This solution has been shown
to agree in part with the results of ``single-angle'' simulations of
supernova neutrino oscillations \cite{Duan:2007fw}. 
These single-angle simulations essentially neglect the anisotropic
nature of the supernova environment by assuming that the flavour
evolution histories of neutrinos along all trajectories are
identical to those along a ``representative''
trajectory, usually taken to be the radial trajectory
\cite{Qian:1994wh,Dasgupta:2008cu}.
The adiabatic precession
solution requires that at any time all neutrinos reside in a pure collective
oscillation mode, the ``precession'' mode, which, as shown
by Duan \etal \cite{Duan:2006an,Duan:2007mv}, would explain the
spectral swap phenomenon in the single-angle simulations.

However, the real supernova environment is highly inhomogeneous and
anisotropic. Here by ``inhomogeneous'' and
``anisotropic'' we are referring to the neutrino fields. Of course,
the matter density distributions in the supernova environment are
also likely to be inhomogeneous and anisotropic.
To date there are a few
``multi-angle'' simulations
\cite{Duan:2006jv,Duan:2006an,Duan:2007bt,Fogli:2007bk} which, like
the single-angle calculations, also adopt spherically symmetric
supernova models but do treat
flavour evolution along different neutrino trajectories in a
self-consistent way. These multi-angle calculations also exhibit
spectral swaps. It is still not understood how the spectral swap
phenomenon arises in the (anisotropic) multi-angle context.
In fact, some studies seem to suggest that collective neutrino
oscillations in the isotropic and anisotropic environments can be very
different. For example,
collective neutrino oscillations of the
bipolar type can experience 
``kinematic decoherence'' and be disrupted in anisotropic environments
\cite{Raffelt:2007yz,EstebanPretel:2007ec}. 
Additionally, a very large matter background can result in neutrino
oscillation phase differences between different neutrino trajectories
in an anisotropic neutrino gas \cite{Qian:1994wh}, and this effect
recently has been shown to result in suppression of collective
neutrino oscillations \cite{EstebanPretel:2008ni}. 

In this paper we discuss a SU($N_\mathrm{f}$) rotation symmetry in the
neutrino flavour evolution equations, where $N_\mathrm{f}=2$ and 3 for
the two-flavour and three-flavour neutrino mixing schemes, respectively. 
The collective precession mode for neutrino oscillations
 can ensue from this
symmetry, even in inhomogeneous, anisotropic environments. 
This result explains the puzzling observations of the 
spectral swapping phenomenon in both the single-angle and
multi-angle simulations of supernova neutrino oscillations.

The rest of this paper is organized as follows.
In \sref{sec:framework} we lay out the general framework for neutrino
flavour transformation and
discuss the SU($N_\mathrm{f}$) rotation symmetry in the flavour evolution
equations for a dense neutrino gas.
In \sref{sec:ap-sol} we show how the collective precession mode for
neutrino oscillations can arise from this symmetry in various
environments. We also give criteria for when the collection precession
mode can occur.
In \sref{sec:sn} we present a new multi-angle
simulation of supernova neutrino oscillations.  We analyze the
results of this calculation guided by our understanding of
 the collective precession mode.
In \sref{sec:conclusions} we give our conclusions.   

\section{Equations of motion and symmetries%
\label{sec:framework}}

\subsection{Neutrino flavour polarization matrix%
\label{sec:eom}}

We are interested in collective flavour oscillations
in neutrino gases in which neutrinos may experience only forward
scattering on other particles (including other neutrinos), 
but where no inelastic scattering occurs. 
When physical conditions change only slowly with spatial dimension,
the flavour content
of neutrinos can be described by semi-classical 
matrices of densities \cite{Sigl:1992fn,Strack:2005ux}
\begin{eqnarray}
\left[\rho_\bfp(t,\bfx)\right]_{\alpha\beta}&= \sum n_{\nu,\bfp}(t,\bfx)
\langle\nu_\alpha|\psi_{\nu,\bfp}(t,\bfx)\rangle
\langle\psi_{\nu,\bfp}(t,\bfx)|\nu_\beta\rangle,
\label{eq:rho-nu}\\
\left[\bar\rho_\bfp(t,\bfx)\right]_{\alpha\beta}&= \sum n_{\bar\nu,\bfp}(t,\bfx)
\langle\bar\nu_\beta|\psi_{\bar\nu,\bfp}(t,\bfx)\rangle
\langle\psi_{\bar\nu,\bfp}(t,\bfx)|\bar\nu_\alpha\rangle,
\label{eq:rho-anu}
\end{eqnarray}
where $\alpha$ and $\beta$ are flavour labels ($e$,
$\mu$, $\tau$),  
$|\psi_{\nu(\bar\nu),\bfp}(t,\bfx)\rangle$ is the state of a
neutrino $\nu$ (antineutrino $\bar\nu$)
with momentum $\bfp$ at time $t$ and position $\bfx$, $n_{\nu(\bar\nu),\bfp}$
is the corresponding neutrino number density, and the summation runs over all
neutrino (antineutrino) states. 
Matrices of densities defined in
\eref{eq:rho-nu} and \eref{eq:rho-anu} 
contain two separate pieces of information. One is the
overall number density of neutrinos or antineutrinos with momentum
$\bfp$ at spacetime point $(t,\bfx)$:
\begin{equation}
n_p(t,\bfx)\equiv\left\{\begin{array}{ll}
\Tr\rho_\bfp(t,\bfx)
&\textrm{ if } p^0 > 0,\\
\Tr\bar\rho_\bfp(t,\bfx)
&\textrm{ if } p^0 < 0.
\end{array}\right.
\label{eq:n}
\end{equation}
Here for compactness we use 4-component vector
$p\equiv[p^0,\bfp]$
to denote a neutrino or antineutrino momentum mode, where $p^0=|\bfp|$
for the neutrino and $-|\bfp|$ for the antineutrino.
Because neutrinos may experience only forward scattering, $n_p(t,\bfx)$
satisfies the conservation equation
\begin{equation}
(\partial_t +\bhp\cdot\bnab) n_p(t,\bfx)=0,
\label{eq:n-cons}
\end{equation}
where $\bhp\equiv\bfp/|\bfp|$ is the unit vector
along the neutrino propagation direction.

The other piece of information contained in the matrix of density is the
``flavour polarization'' of the neutrino. This is in analogy to, e.g., the spin
polarization of an electron gas. We define
a traceless ``neutrino flavour polarization matrix'', or ``polarization
matrix'' for short, 
\begin{equation}
\sfP_p(t,\bfx)\equiv\left\{\begin{array}{ll}
n_p^{-1}(t,\bfx)\rho_p(t,\bfx)-N_\mathrm{f}^{-1}\mathsf{I}&
\textrm{ if } p^0>0,\\
-n_p^{-1}(t,\bfx)\rho_p(t,\bfx)+N_\mathrm{f}^{-1}\mathsf{I}&
\textrm{ if } p^0<0,
\end{array}\right.
\label{eq:P}
\end{equation}
where $\mathsf{I}$ is the identity matrix in flavour space, and
$N_\mathrm{f}=2$ and 3 for two-flavour and three-flavour mixing
schemes, respectively.
In \eref{eq:P} we define the polarization matrices for neutrinos and
antineutrinos with opposite signs, with the understanding that
antiparticles are ``negative particles'' or ``holes'' in the particle
sea. This sign convention
will make the equations of motion (e.o.m.) more succinct 
and is especially appropriate in the two-flavour mixing scheme where
$\mathbf{2}$ and $\bar{\mathbf{2}}$,
the fundamental representations of the SU(2) group, are equivalent
\cite{Duan:2005cp} (see \sref{sec:NFIS}).
The e.o.m.~for polarization matrix $\sfP_p(t,\bfx)$ can be 
derived easily from that for $\rho_p(t,\bfx)$
\cite{Sigl:1992fn,Strack:2005ux,Cardall:2007zw} and is
\begin{equation}
(\partial_t+\bhp\cdot\bnab)\sfP_p(t,\bfx)
=-\rmi[\sfH_p(t,\bfx),\sfP_p(t,\bfx)].
\label{eq:eom-P}
\end{equation}
Throughout this paper we assume a vanishing $CP$-violating phase.
(See \cite{Gava:2008rp} for a discussion of collective neutrino
oscillations with a non-vanishing $CP$-violating phase.)

The Hamiltonian for polarization matrix $\sfP_p(t,\bfx)$ is
\begin{eqnarray}
\sfH_p(t,\bfx)
&=\sfH_{p^0}^\ext(t,\bfx)
+\sfH_{\bhp}^{\nu\nu}(n_{p^\prime}(t,\bfx),\sfP_{p^\prime}(t,\bfx)|\forall
p^\prime),
\nonumber\\
&=\sfH_{p^0}^\vac+\sfH^\matt(t,\bfx)
+\sfH_{\bhp}^{\nu\nu}(n_{p^\prime}(t,\bfx),\sfP_{p^\prime}(t,\bfx)|\forall
p^\prime).
\label{eq:H}
\end{eqnarray}
The background ``neutrino field''
\begin{equation}
\fl
\sfH_{\bhp}^{\nu\nu}(n_{p^\prime}(t,\bfx),\sfP_{p^\prime}(t,\bfx)|\forall
p^\prime)
=\sqrt{2}\GF\sum_{p^\prime} 
(1-\bhp\cdot\bhp^\prime)n_{p^\prime}(t,\bfx)\sfP_{p^\prime}(t,\bfx)
\label{eq:Hnunu}
\end{equation} 
is a function of both neutrino number densities $n_{p^\prime}(t,\bfx)$
and neutrino flavour polarization matrices $\sfP_{p^\prime}(t,\bfx)$,
and does not depend on the energy of the test
neutrino. For convenience we will drop the symbol ``$\forall
p^\prime$'' with the understanding that $p^\prime$ in
$\sfH_{\bhp}^{\nu\nu}(n_{p^\prime}(t,\bfx),\sfP_{p^\prime}(t,\bfx))$
refers to all neutrino modes.
In \eref{eq:Hnunu} we use $\sum_{p^\prime}$ to denote the integration over
$(p^{\prime})^0$ and $\bhp^\prime$:
\begin{equation}
\sum_{p^\prime}\equiv\int_{-\infty}^\infty\rmd(p^{\prime})^0\int\!\rmd\bhp^\prime.
\label{eq:sum-p}
\end{equation}
The procedure implied in equation \eref{eq:sum-p} is tantamount to a sum over
neutrino and antineutrino energies and trajectory directions.
The ``vacuum field''
\begin{equation}
\sfH_{p^0}^\vac=\frac{\mathsf{M}^2}{2p^0}
\label{eq:Hvac}
\end{equation}
generates vacuum oscillations, where $\mathsf{M}$ is the
neutrino mass matrix.
Because the trace of a Hamiltonian has no
effect on neutrino oscillations, we will take $\sfH_{p^0}^\vac$ to be
traceless hereafter. With this convention \eref{eq:Hvac} 
in the vacuum mass basis becomes
\begin{equation}
\sfH_{p^0}^\vac=-\frac{\dm_{21}}{2p^0}\frac{\Lambda_3}{2}
-\left(\frac{\dm_{31}+\dm_{32}}{4p^0}\right)
\frac{\Lambda_8}{\sqrt{3}},
\label{eq:HV-3flavour}
\end{equation}
where $\dm_{ij}=m_i^2-m_j^2$ is the difference between the squares of
the mass eigenvalues corresponding to mass
eigenstates $|\nu_i\rangle$ and $|\nu_j\rangle,$ and 
$\Lambda_a$ ($a=1,2,\ldots,8$) are the Gell-Mann matrices.
For the two-flavour mixing scheme
\begin{equation}
\sfH_{p^0}^\vac=-\frac{\dm}{2p^0}\frac{\Lambda_3}{2},
\label{eq:HV-2flavour}
\end{equation}
where $\Lambda_3$ is the third Pauli matrix. The $2\times2$ case is
analogous to the $3\times3$ case discussed above except 
 $\Lambda_a$ ($a=1,2,3$) are the Pauli matrices.
The ``matter field'' in the \textit{flavour basis} in
the supernova environment is
\begin{equation}
\sfH^\matt(t,\bfx)=\lambda(t,\bfx)\,\diag[1,0,0]
=\sqrt{2}\GF n_e(t,\bfx)\,\diag[1,0,0],
\end{equation}
where $\GF$ is the Fermi constant, and $n_e(t,\bfx)$ is the net electron
number density.
The vacuum field $\sfH_{p^0}^\vac$ and the matter field
$\sfH^\matt(t,\bfx)$ together constitute the total external field,
$\sfH^\ext_{p^0}(t,\bfx)$, which does not depend on neutrino flavours.

\subsection{Vector representation of the polarization matrix%
\label{sec:NFIS}}

An $N_\mathrm{f}\times N_\mathrm{f}$, traceless, Hermitian matrix
$\mathsf{A}$ can be written in terms of vector $\vec{A}$ as
\begin{equation}
\mathsf{A}=\frac{\vec\Lambda}{2}\cdot\vec{A}
\equiv\sum_a \frac{\Lambda_a}{2} A_a.
\label{eq:vP}
\end{equation}
(We have adopted the convention in
\cite{Dasgupta:2008cu}. We use bold-faced letters, e.g.,
$\bi{A}$ to denote 3-vectors in coordinate space, sans-serif letters,
e.g., $\mathsf{A}$ for matrices in flavour space, and letters 
with an arrow, e.g., $\vec{A}$ for vectors in flavour space.)
The e.o.m.\ for the ``polarization vector'' $\vP_p(t,\bfx)$ is
\cite{Dasgupta:2007ws} 
\begin{equation}
(\partial_t+\bhp\cdot\bnab) \vP_p(t,\bfx)
=\vH_p(t,\bfx)\times\vP_p(t,\bfx),
\label{eq:eom-vP}
\end{equation}
where the cross product between two vectors is defined by the structure
constants $f_{abc}$ of the SU($N_\mathrm{f}$) group \cite{Kim:1987bv}:
\begin{equation}
(\vec{A}\times\vec{B})_a\equiv f_{abc}A_b B_c.
\end{equation}

The definition of the polarization vector $\vP_p$ 
given by \eref{eq:P} and \eref{eq:vP}
for the
antineutrino has a different sign as compared with that in
\cite{Sigl:1992fn} and with the 
8-dimensional Bloch vector in \cite{Dasgupta:2007ws}. 
In addition to making the expression of $\vH^{\nu\nu}_{\bhp}(t,\bfx)$ [see
  \eref{eq:Hnunu}] more compact, this convention is 
especially convenient in the two-flavour mixing scheme. In studying
collective neutrino oscillations 
the corotating-frame transformation technique \cite{Duan:2005cp}
is frequently used. 
A corotating-frame transformation corresponds to rewriting \eref{eq:eom-vP}
in a reference frame that rotates about $\ve_3$ with
angular frequency $\omega_0$, where $\ve_3$ is the unit basis vector
corresponding to $\Lambda_3$ in the vacuum mass basis. 
According to \eref{eq:H} and \eref{eq:Hvac}
this transformation is equivalent to a change in the momentum of the
neutrino $p\rightarrow p^\prime$ where 
\begin{equation}
\frac{\delta m^2}{2(p^\prime)^0}
=\frac{\delta m^2}{2p^0}-\omega_0
\quad\textrm{and}\quad
\bhp^\prime=\bhp.
\label{eq:p-shift}
\end{equation}
With the traditional definition,
the direction of the polarization vector $\vP_p$ must be reversed
when $p^0$ changes sign under the transformation
\eref{eq:p-shift}. The polarization vector defined by \eref{eq:P} and
\eref{eq:vP}, however, is invariant under such transformations.
This definition has already been adopted in some recent literature,
e.g., \cite{Raffelt:2008hr}. 

We can define the magnitude of $\sfP_p$ as 
\begin{equation}
|\sfP_p|\equiv|\vP_p|\equiv\sqrt{\sum_a P_a^2} 
\leq\sqrt{\frac{2}{N_\mathrm{f}}(N_\mathrm{f}-1)}.  
\end{equation}
The equal sign in the above strict inequality relation applies only if
the neutrino state is a pure (quantum) state, i.e., can be
described by a single ket. 
In forward-scattering the coherence of the neutrino is
not lost and, therefore, $|\sfP_p(t,\bfx)|$ also obeys the conservation
equation
\begin{equation}
(\partial_t+\bhp\cdot\bnab)|\sfP_p(t,\bfx)|=0.
\end{equation}

In the two-flavour mixing scheme a notation related to the polarization
vector is the neutrino flavour 
isospin (NFIS) \cite{Duan:2005cp}. It can be defined as
\begin{equation}
\vs_p(t,\bfx)\equiv \frac{\vP(t,\bfx)}{2|\vP(t,\bfx)|}
\label{eq:P-s-pure}
\end{equation}
if the neutrino state is a pure state. It obeys the e.o.m.
\begin{eqnarray}
\fl
(\partial_t+\bhp\cdot\bnab)\vs_p(t,\bfx)
&=\vs_p(t,\bfx)\times[-\vH^\vac_{p^0}-\vH^\matt(t,\bfx)
\nonumber\\
&\quad-2\sqrt{2}\GF\sum_{p^\prime}(1-\bhp\cdot\bhp^\prime)
n_{p^\prime}(t,\bfx)\vs_{p^\prime}(t,\bfx)].
\label{eq:eom-s}
\end{eqnarray}
\Eref{eq:eom-s} shows that two NFIS's $\vs_p(t,\bfx)$ and
$\vs_{p^\prime}(t,\bfx)$ at the same space-time point are
``antiferromagnetically'' coupled.
Note that we have defined Hamiltonian vector fields in the same way as
\cite{Raffelt:2007cb} but differing by a minus sign from those in the
original NFIS notation \cite{Duan:2005cp}.
In supernovae the neutrinos in a given momentum mode usually are not
in a pure state. 
Assuming that supernova neutrinos are emitted in pure flavour states
at the neutrino sphere and subsequently encounter only forward-scattering,
we can write
\begin{equation}
n_p(t,\bfx)\vP_p(t,\bfx)=
2\sum_\alpha n_{\alpha,p}(t,\bfx)\vs_{\alpha,p}(t,\bfx).
\label{eq:P-s}
\end{equation}
In \eref{eq:P-s} $\vs_{\alpha,p}(t,\bfx)$ is the NFIS
that represents the flavour state of the neutrino or antineutrino at $(t,\bfx)$
which is pure $\nu_\alpha$ or $\bar\nu_\alpha$ at the neutrino sphere,
and $n_{\alpha,p}(t,\bfx)$ is the associated
neutrino number density. 
The corresponding replacements in \eref{eq:eom-s} are
\[\vs_p\rightarrow\vs_{\alpha,p},\quad
\vs_{p^\prime}\rightarrow\vs_{\alpha^\prime,p^\prime},\quad
n_{p^\prime}\rightarrow n_{\alpha^\prime,p^\prime}\quad
\textrm{and}\quad
\sum_{p^\prime}\rightarrow\sum_{\alpha^\prime,p^\prime}.\]
Equivalently, we can insist on the definition \eref{eq:P-s-pure} and
replace $n_p(t,\bfx)$ in \eref{eq:eom-s} by 
\begin{equation}
n_p^\prime(t,\bfx)\equiv n_p(t,\bfx)|\vP(t,\bfx)|.
\end{equation}
In the latter approach, NFIS $\vs_p(t,\bfx)$ represents the ``average
flavour state'' of the neutrino, and $n_p^\prime(t,\bfx)$ is the ``net
number density'' of the neutrino in the flavour state 
represented by $\vs_p(t,\bfx)$.

\subsection{Symmetries and conservation laws%
\label{sec:L-cons}}

From \eref{eq:Hnunu} it is easy to show that
$\sfH_{\bhp}^{\nu\nu}(n_{p^\prime}(t,\bfx),\sfP_{p^\prime}(t,\bfx))$
satisfies two important identities:
\begin{equation}
\sum_p n_p(t,\bfx)[\sfP_p(t,\bfx),
\sfH_{\bhp}^{\nu\nu}(n_{p^\prime}(t,\bfx),\sfP_{p^\prime}(t,\bfx))]=0;
\label{eq:Hnunu-sym}
\end{equation}
and
\begin{equation}
\fl
\sfU(t,\bfx)
\sfH_{\bhp}^{\nu\nu}(n_{p^\prime}(t,\bfx),\sfP_{p^\prime}(t,\bfx))
\sfU^\dagger(t,\bfx)
=
\sfH_{\bhp}^{\nu\nu}(n_{p^\prime}(t,\bfx),
\sfU(t,\bfx)\sfP_{p^\prime}(t,\bfx)\sfU^\dagger(t,\bfx)),
\label{eq:Hnunu-func}
\end{equation}
where $\sfU(t,\bfx)$ is an arbitrary unitary matrix.

\Eref{eq:Hnunu-sym} implies that, if 
\begin{equation}
[\sfG, \sfH_{p^0}^\ext(t,\bfx)] = 0,
\label{eq:G-Hext}
\end{equation}
where $\sfG$ is a constant $N_\mathrm{f}\times N_\mathrm{f}$
traceless  Hermitian matrix, then the lepton current $L^\mu(t,\bfx)$
with temporal and spatial components
\begin{eqnarray}
L^0(t,\bfx) &\equiv
\sum_p n_p(t,\bfx) \Tr\left[\sfP_p(t,\bfx)\,\sfG\right],\\
\bi{L}(t,\bfx) &\equiv
\sum_p \bhp\, n_p(t,\bfx) \Tr\left[\sfP_p(t,\bfx)\,\sfG\right]
\end{eqnarray}
satisfies the continuity equation
\begin{equation}
\partial_\mu L^\mu(t,\bfx)=\partial_t L^0(t,\bfx)+\bnab\cdot\bi{L}(t,\bfx) = 0.
\label{eq:cons-law}
\end{equation}
This can be easily shown using
\eref{eq:n-cons}, \eref{eq:eom-P}, \eref{eq:H}, \eref{eq:Hnunu-sym}
and \eref{eq:G-Hext}: 
\begin{eqnarray}
\fl
\partial_\mu L^\mu(t,\bfx)
&=\rmi\sum_p n_p(t,\bfx)\Tr\left(
[\sfP_p(t,\bfx),\sfH_p(t,\bfx)]\,\sfG\right)
\nonumber\\
\fl
&= \rmi\sum_p n_p(t,\bfx)\Tr\left(
[\sfP_p(t,\bfx),\sfH_{p^0}^\ext(t,\bfx)]\,\sfG\right)
\nonumber\\
\fl
&= \rmi\sum_p n_p(t,\bfx)\Tr\left(\sfP_p(t,\bfx)\,
[\sfH_{p^0}^\ext(t,\bfx),\sfG]\right) = 0.
\label{eq:G-trace}
\end{eqnarray}

\Eref{eq:Hnunu-func} implies that, if \eref{eq:G-Hext} is true, then
the e.o.m.\ \eref{eq:eom-P} for the polarization matrix is invariant
under the global (i.e., independent of $p$,
$t$ and $\bfx$) SU($N_\mathrm{f}$) transformation
\begin{equation}
\sfP_p(t,\bfx)\longrightarrow\stP_p(t,\bfx)\equiv
\exp(-\rmi\phi\sfG)\sfP_p(t,\bfx)\exp(\rmi\phi\sfG),
\label{eq:G-transform}
\end{equation}
where $\phi$ is an arbitrary constant scalar.
This is because \eref{eq:G-Hext} implies
\begin{equation}
\exp(-\rmi\phi\sfG) \sfH_{p^0}^\ext(t,\bfx)
\exp(\rmi\phi\sfG) = \sfH_{p^0}^\ext(t,\bfx).
\label{eq:Hext-G-transform}
\end{equation}
Using \eref{eq:eom-P},
\eref{eq:Hnunu-func} and \eref{eq:Hext-G-transform} it can be shown that
\begin{equation}
\fl
(\partial_t + \bhp\cdot\bnab) \stP_p(t,\bfx)
=-\rmi\left[\sfH^\ext_{p^0}(t,\bfx)
+\sfH^{\nu\nu}_{\bhp}(n_{p^\prime}(t,\bfx),\stP_{p^\prime}(t,\bfx)),\,
\stP_p(t,\bfx)\right].
\label{eq:eom-P-G-transform}
\end{equation}

In the two-flavour mixing scheme it is obvious that $\Lambda_3$ (in the
vacuum mass basis) commutes with $\sfH_{p^0}^\ext=\sfH_{p^0}^\vac$
in \eref{eq:HV-2flavour} in the absence of ordinary matter. According
to the above discussion, the 
e.o.m.\ \eref{eq:eom-vP} for all polarization vectors
$\vP_p(t,\bfx)$ are invariant under simultaneous rotations about
$\ve_3$. 
In a homogeneous and isotropic neutrino gas a
collective neutrino oscillation mode, which is represented by the collective
precession of all polarization vectors about $\ve_3$, can arise
because of this 
symmetry \cite{Duan:2007mv,Duan:2007fw}. 
This precession mode of collective neutrino oscillations
will ultimately cause the energy spectra of neutrinos with different
flavours to be swapped at a critical energy $\Es$, a phenomenon known as
the ``stepwise spectral swapping'' \cite{Duan:2006an}. Not
surprisingly, the value of $\Es$ is determined by the conserved lepton
number $L^0$ associated with $\Lambda_3$
\cite{Hannestad:2006nj,Raffelt:2007cb}. 
Similar conclusions have also been drawn for homogeneous, isotropic
neutrino gases in the three-flavour mixing scheme
\cite{Duan:2008za,Dasgupta:2008cd}.  
 
Approximate symmetries and conservation laws can exist for scenarios
where matter densities are large. Noting that $\sfH^\matt(t,\bfx)$ is
invariant under any rotation in the $\nu_\mu$--$\nu_\tau$ subspace,
one can diagonalize the $\nu_\mu$--$\nu_\tau$ submatrix of
$\sfH^\ext_{p^0}(t,\bfx)$ by a rotation
$(\nu_e,\nu_\mu,\nu_\tau)\rightarrow(\nu_e,\nu_{\mu^\prime},\nu_{\tau^\prime})$. In
this new basis the external field is written as
\begin{eqnarray}
\sfH_{p^0}^\ext(t,\bfx)&=\frac{1}{2p^0}\left[\begin{array}{lll}
m_{ee}^2+2\sqrt{2}p^0\GF n_e(t,\bfx) & m_{e\mu^\prime}^2 & m_{e\tau^\prime}^2 \\
m_{e\mu^\prime}^2 & m_{\mu^\prime\mu^\prime}^2 & 0 \\
m_{e\tau^\prime}^2 & 0 & m_{\tau^\prime\tau^\prime}^2
\end{array}\right]
\nonumber\\
&\simeq  \frac{1}{2p^0}\diag[m_{ee}^2+2\sqrt{2}p^0\GF n_e(t,\bfx),\,
m_{\mu^\prime\mu^\prime}^2,\, m_{\tau^\prime\tau^\prime}^2],
\label{eq:Hext-diag}
\end{eqnarray} 
which is approximately diagonalized for any neutrino mode $p$ if 
$\GF n_e(t,\bfx)\gg|\dm_{ij}/(2p^0)|$ (e.g., \cite{Dighe:1999bi}).
The approximate symmetries of the neutrino system about $\Lambda_3$
and $\Lambda_8$, therefore, exist in the new basis
$(\nu_e,\nu_{\mu^\prime},\nu_{\tau^\prime})$ instead of the vacuum mass
basis \cite{Duan:2008za}.

\section{Collective precession mode for neutrino oscillations%
\label{sec:ap-sol}}

In this section we discuss the collective precession mode in the
two-flavour mixing scheme. This collective mode solution can arise
in various environments because of the symmetry discussed in
\sref{sec:L-cons}. Generalization to the full
three-flavour mixing scheme is straightforward when the polarization matrix
representation is used \cite{Duan:2008za}. 

\subsection{Stationary, homogeneous and isotropic environments%
\label{sec:stationary-isotropic}}

First we shall use the symmetry viewpoint to discuss
neutrino oscillations in the collective precession mode in
 stationary, homogeneous, isotropic environments.
In such environments no physical quantity depends on
the neutrino propagation direction $\bhp$ and
neither the external field  
$\sfH_{p^0}^\ext$ nor the neutrino number density $n_{p^0}$  varies
with space or time. 
Also in this case, the polarization matrices
$\sfP_{p^0}(t)$ are uniform and isotropic, which means the neutrino
self-interaction potential
\begin{equation}
\sfH^{\nu\nu}(n_{(p^\prime)^0},\sfP_{(p^\prime)^0}(t))
=\sqrt{2}\GF\sum_{(p^\prime)^0} 
n_{(p^\prime)^0}\sfP_{(p^\prime)^0}(t)
\label{eq:Hnunu-homo}
\end{equation}
does not depend on the momentum of the test neutrino.
Suppose that the set of variables $\{\Omega,\stP_{p^0}|\forall p^0\}$ solve
the following equations
\begin{eqnarray}
\left[\sfH_{p^0}^\ext +
  \sfH^{\nu\nu}(n_{(p^\prime)^0},\stP_{(p^\prime)^0}) 
+ \Omega\frac{\Lambda_3}{2},
\stP_{p^0}\right] = 0,
\label{eq:prec-ansatz-homo}\\
\sum_{p^0} n_{p^0}
\Tr\left(\stP_{p^0}\Lambda_3\right)
= L^0,
\label{eq:L-cons-homo}
\end{eqnarray}
where $\Omega$ is a scalar independent of $p^0$, $\stP_{p^0}$ are
traceless Hermitian matrices, and 
$L^0$ is a constant. 
Here we adopt the the vacuum mass basis $(\nu_1,\nu_2)$
if $\sfH^\matt(t)$ is negligible, and the flavour basis
$(\nu_e,\nu_{\mu^\prime})$ (which is also the matter basis)  if
$\sfH^\matt(t)$ is dominant. Therefore,
\begin{equation}
\fl
\sfH^\ext_{p^0} \simeq -(\omega_{p^0}-\lambda)\frac{\Lambda_3}{2} 
\simeq \left\{\begin{array}{ll}
-\frac{\dm}{2p^0}\frac{\Lambda_3}{2}
&\textrm{ if } \lambda \textrm{ is negligible,}\\
-\left(\frac{\Delta m^{\prime2}}{2p^0}-\lambda\right)\frac{\Lambda_3}{2}
&\textrm{ if } \lambda \textrm{ is very large,}
\end{array}\right.
\label{eq:Hext-homo}
\end{equation}
where $\Delta m^{\prime2}=m_{\mu^\prime\mu^\prime}^2-m_{ee}^2$ [see
  \eref{eq:Hext-diag}]. 
In either case we have
\begin{equation}
[\sfH_{p^0}^\ext,\,\Lambda_3] = 0.
\label{eq:Hext-Lambda3}
\end{equation}

Using \eref{eq:Hnunu-func}, \eref{eq:prec-ansatz-homo} and
\eref{eq:Hext-Lambda3} we can  show easily that 
\begin{equation}
\sfP_{p^0}(t)=\exp\left[\rmi(\Omega t-\phi_0)\frac{\Lambda_3}{2}\right]
\stP_{p^0}\exp\left[-\rmi(\Omega t-\phi_0)\frac{\Lambda_3}{2}\right]
\label{eq:prec-sol-homo}
\end{equation}
is a solution to the e.o.m.\ [\eref{eq:eom-P} without spatial dependence]: 
\begin{eqnarray}
\frac{\rmd}{\rmd t} \sfP_{p^0}(t)
&= \rmi\left[\Omega\frac{\Lambda_3}{2},\,\sfP_{p^0}(t)\right]
\nonumber\\
&=
-\rmi\left[\sfH_{p^0}^\ext+
\sfH^{\nu\nu}(n_{(p^\prime)^0},\sfP_{(p^\prime)^0}(t)),\, 
\sfP_{p^0}(t)\right],
\end{eqnarray}
where $\phi_0$ is a constant.
The solution obtained from \eref{eq:prec-ansatz-homo},
\eref{eq:L-cons-homo} and \eref{eq:prec-sol-homo}
is called the ``precession solution''.

\Eref{eq:prec-ansatz-homo} is effectively a set of $3\times N_\mathrm{en}$
coupled nonlinear integral equations, where
\begin{equation}
N_\mathrm{en}\equiv\sum_{p^0} 1
\end{equation}
is the number of neutrino/antineutrino energy modes.
Raffelt and Smirnov \cite{Raffelt:2007cb} pointed out that
\eref{eq:prec-ansatz-homo} can be reduced to two nonlinear integral
equations. This can be shown as follows. We define
\begin{equation}
\stH_{p^0}
\equiv \sfH_{p^0}^\ext +
  \sfH^{\nu\nu}(n_{(p^\prime)^0},\stP_{(p^\prime)^0}) 
+\Omega\frac{\Lambda_3}{2}.
\end{equation}
This Hamiltonian $\stH_{p^0}$ 
commutes with $\stP_{p^0}$ if \eref{eq:prec-ansatz-homo} is
true. This means that $\vtP_{p^0}$ 
is either aligned or antialigned with the vector field $\vtH_{p^0}$:
\begin{equation}
\vtP_{p^0} = 
\frac{\epsilon_{p^0}|\vtP_{p^0}|}{|\vtH_{p^0}|}
\vtH_{p^0},
\label{eq:alignment-homo}
\end{equation}
where $\epsilon_{p^0}=+1$ ($-1$) if $\vtP_{p^0}$ is aligned
(antialigned) with $\vtH_{p^0}$.

Using \eref{eq:Hnunu-homo}, \eref{eq:Hext-homo}, and
\eref{eq:alignment-homo} one can obtain \cite{Raffelt:2007cb}
\begin{eqnarray}
\sum_{p^0} 
\frac{\epsilon_{p^0}\, n_{p^0} |\vtP_{p^0}|}%
{\sqrt{[(\omega_{p^0}-\lambda-\Omega)/\mu(n_\nu^\tot) - 
\langle \tP_3\rangle]^2 + \langle \tP_1\rangle^2}}
&=n_\nu^\tot,
\label{eq:unitary-cond-homo}\\
\sum_{p^0} 
\frac{\epsilon_{p^0}\, n_{p^0}\, \omega_{p^0} |\vtP_{p^0}|}%
{\sqrt{[(\omega_{p^0}-\lambda-\Omega)/\mu(n_\nu^\tot) - 
\langle \tP_3\rangle]^2 + \langle \tP_1\rangle^2}}
&=(\lambda+\Omega)n_\nu^\tot,
\label{eq:omega-cond-homo}
\end{eqnarray}
where 
\begin{equation}
n_\nu^\tot \equiv \sum_{p^0}
n_{p^0}
\end{equation}
is the total neutrino number density,
\begin{equation}
\mu(n_\nu^\tot)\equiv \sqrt{2}\GF n_\nu^\tot,
\end{equation}
and
\begin{equation}
\langle \vtP\rangle\equiv
\frac{1}{n_\nu^\tot}\sum_{p^0} n_{p^0} \vtP_{p^0}
\end{equation}
is the average polarization vector.
Note that we have chosen an appropriate value of $\phi_0$ so that
$\langle\tP_2\rangle=0$. 
Given the set of parameters
$\{\lambda,|\vP_{p^0}|=|\vtP_{p^0}|,n_{p^0},\epsilon_{p^0}|\forall p^0\}$,
 \eref{eq:L-cons-homo},
\eref{eq:unitary-cond-homo} and \eref{eq:omega-cond-homo} can be
solved for the set of quantities
$\{\Omega,\langle \tP_1\rangle,\langle \tP_3\rangle\}$ 
which, in turn,
determine $\vtP_{p^0}$ through \eref{eq:alignment-homo}.

It is obvious from \eref{eq:unitary-cond-homo} and
\eref{eq:omega-cond-homo} that $\Omega$  has a simple
dependence on $\lambda$:
\begin{equation}
\Omega(\lambda) = \Omega|_{\lambda=0} - \lambda.
\label{eq:Omega-lambda}
\end{equation} 
In other words, in the presence of a large matter density, 
$\Omega|_{\lambda=0}$ can be calculated as if there is no ordinary
matter (but with $\omega_{p^0}=\Delta m^{\prime 2}/2p^0$), and then 
$\Omega$ can be obtained using \eref{eq:Omega-lambda}. We note that 
$\langle\vtP\rangle$ is independent of $\lambda$ and, therefore, in
this case
neutrino flavour transformation does not depend on the matter density
except for an extra rotation in \eref{eq:prec-sol-homo}.  This result is
expected using the corotating-frame technique \cite{Duan:2005cp}.

We note that \eref{eq:prec-ansatz-homo} [or
  \eref{eq:alignment-homo}--\eref{eq:omega-cond-homo}] and
\eref{eq:L-cons-homo} are \emph{not} guaranteed to have a solution or
solutions. However, if one can solve these equations, then a
``precession solution'' \eref{eq:prec-sol-homo} automatically obtains because of
\eref{eq:Hext-Lambda3}.
This precession solution corresponds to
the collective precession of polarization vectors about the $\ve_3$ axis.

\subsection{Slowly-varying, homogeneous and isotropic environments%
\label{sec:adiabatic-isotropic}}

Once the collective precession mode discussed in
\sref{sec:stationary-isotropic} is established in the homogeneous, isotropic
neutrino gas, it can be expected that, as $\lambda(t)$ and $n_\nu^\tot(t)$
vary slowly with time $t$ (but with $n_{p^0}(t)/n_\nu^\tot(t)$ fixed
for all $p^0$), the collective mode continues and transforms
adiabatically. In other words, we still have
\begin{equation}
\sfP_{p^0}(t)=\exp\left[-\rmi\phi(t)\frac{\Lambda_3}{2}\right]
\stP_{p^0}(t)\exp\left[\rmi\phi(t)\frac{\Lambda_3}{2}\right],
\label{eq:prec-sol-homo-var}
\end{equation}
except that $\stP_{p^0}(t)=\stP_{p^0}(n_\nu^\tot(t))$ has a weak dependence on $t$
through $n_\nu^\tot(t)$. Because neutrinos encounter only
forward-scatterings and, therefore, $n_{p^0}(t)/n_\nu^\tot(t)$ is
constant, we have
\begin{equation}
\frac{\rmd}{\rmd t}\langle\tP_3(t)\rangle
=\frac{\rmd}{\rmd t}\left[\frac{1}{n_\nu^\tot(t)}
\sum_{p^0}n_{p^0}(t)\Tr\left(\sfP_{p^0}(t)\,\Lambda_3\right)\right] = 0.
\label{eq:L-cons-homo-var}
\end{equation}
Similar to \eref{eq:unitary-cond-homo} and \eref{eq:omega-cond-homo}
we have
\begin{eqnarray}
\fl
\sum_{p^0} \frac{\epsilon_{p^0}\, n_{p^0}(t)\, |\vtP_{p^0}|}%
{\sqrt{\{[\omega_{p^0}-\lambda(t)+\phi'(t)]/\mu(t) - 
\langle \tP_3(t)\rangle\}^2 + \langle
\tP_1(t)\rangle^2}}
=n_\nu^\tot(t),
\label{eq:unitary-cond-homo-var}\\
\fl
\sum_{p^0} 
\frac{\epsilon_{p^0}\, n_{p^0}(t)\, \omega_{p^0} |\vtP_{p^0}|}%
{\sqrt{\{[\omega_{p^0}-\lambda(t)+\phi'(t)]/\mu(t) - 
\langle \tP_3(t)\rangle\}^2 + \langle \tP_1(t)\rangle^2}}
=[\lambda(t)-\phi'(t)]n_\nu^\tot(t),
\label{eq:omega-cond-homo-var}
\end{eqnarray}
where $\mu(t)=\mu(n_\nu^\tot(t))$ depends on $t$ through $n_\nu^\tot(t)$.
In \eref{eq:unitary-cond-homo-var} and
\eref{eq:omega-cond-homo-var} $\epsilon_{p^0}$ is constant for the
adiabatic process, and $|\vtP_{p^0}|$ does not vary with time because
the coherence in neutrino mixing is maintained.

Using \eref{eq:L-cons-homo-var}--\eref{eq:omega-cond-homo-var} 
 $\langle\vtP(t)\rangle$ and $\phi(t)$ can be found. 
The polarization vectors $\vtP_{p^0}(t)$ can then be found by using
\begin{equation}
\vtP_{p^0}(t) = 
\frac{\epsilon_{p^0}|\vtP_{p^0}|}{|\vtH_{p^0}(t)|}
\vtH_{p^0}(t),
\label{eq:alignment-homo-var}
\end{equation}
where
\begin{equation}
\vtH_{p^0}(t)
= \vH^\ext_{p^0}(\lambda(t))
+\mu(t)\langle\vtP(t)\rangle
-\phi'(t)\frac{\Lambda_3}{2}.
\end{equation}
The full adiabatic precession solution is then obtained using
\eref{eq:prec-sol-homo-var}. 

\subsection{Stationary, homogeneous but anisotropic environments%
\label{sec:stationary-anisotropic}}

We now consider a neutrino gas in a stationary, homogeneous
environment. By stationary and homogeneous we mean that neither
$\lambda$ or $n_p$ varies with space or time. We note that the
environment is anisotropic if $n_p$ depends on the neutrino
propagation direction $\bhp$. We also note that $\sfP_p(t,\bfx)$
can vary with time and/or space even if the environment is stationary
and homogeneous. Like in \sref{sec:stationary-isotropic} we assume
that the set of quantities $\{\Omega,\bfK,\stP_p|\forall p\}$ is a solution to 
\begin{eqnarray}
\left[\sfH_{p^0}^\ext +
  \sfH^{\nu\nu}_{\bhp}(n_{p^\prime},\stP_{p^\prime}) 
+ (\Omega-\bhp\cdot\bfK)\frac{\Lambda_3}{2},
\stP_p\right] = 0,
\label{eq:prec-ansatz-anisotropic}\\
\sum_{p} n_p
\Tr\left(\stP_{p}\Lambda_3\right)
= L^0,
\label{eq:L0-cons-anisotropic}\\
\sum_{p} \bhp\, n_p
\Tr\left(\stP_{p}\Lambda_3\right)
= \bi{L},
\label{eq:Lv-cons-anisotropic}
\end{eqnarray}
where $\Omega$ and $L^0$ are constant scalars,
$\bfK$ and $\bi{L}$ are constant vectors,
and $\stP_p$ are traceless Hermitian matrices.
Using \eref{eq:Hnunu-func}, \eref{eq:Hext-Lambda3} and
\eref{eq:prec-ansatz-anisotropic} we can  show easily that
\begin{equation}
\fl
\sfP_p(t,\bfx)=\exp\left[\rmi(\Omega t - \bfK\cdot\bfx-\phi_0)
\frac{\Lambda_3}{2}\right]
\stP_p\exp\left[-\rmi(\Omega t - \bfK\cdot\bfx-\phi_0)
\frac{\Lambda_3}{2}\right]
\label{eq:prec-sol-anisotropic}
\end{equation}
is a solution to the e.o.m.\ \eref{eq:eom-P}:
\begin{eqnarray}
\fl
(\partial_t+\bhp\cdot\bnab) \sfP_p(t,\bfx)
&= \rmi\left[(\Omega- \bhp\cdot\bfK)\frac{\Lambda_3}{2},\,
\sfP_p(t,\bfx)\right]
\nonumber\\
\fl
&=
-\rmi\left[\sfH_{p^0}^\ext+
\sfH_{\bhp}^{\nu\nu}(n_{p^\prime},\sfP_{p^\prime}(t,\bfx)),\, 
\sfP_p(t,\bfx)\right],
\end{eqnarray}
where $\phi_0$ is a constant.

Like in the stationary, homogeneous and isotropic case,
\eref{eq:prec-ansatz-anisotropic}--\eref{eq:Lv-cons-anisotropic} are
not guaranteed to have a solution or solutions. If such a solution
does exist, however, the symmetry in the neutrino flavour evolution
equations automatically gives a collective precession mode solution
for neutrino oscillations \eref{eq:prec-sol-anisotropic}.

\Eref{eq:prec-ansatz-anisotropic} implies that $\vtP_p$ is either
aligned or antialigned with the vector field 
\begin{eqnarray}
\vtH_p &\equiv \vH^\ext_{p^0} 
+ \vH_{\bhp}^{\nu\nu}(n_{p^\prime},\vtP_{p^\prime})
+(\Omega-\bhp\cdot\bfK)\frac{\Lambda_3}{2}
\nonumber\\
&= \vH^\ext_{p^0} 
+\sum_{\bhp^\prime}\mu_{\bhp\cdot\bhp^\prime}(n_\nu^\tot)\langle\vtP_{\bhp}\rangle
+(\Omega-\bhp\cdot\bfK)\frac{\Lambda_3}{2},
\label{eq:vtH-anisotropic}
\end{eqnarray}
and
\begin{equation}
\vtP_p = 
\frac{\epsilon_p|\vtP_p|}{|\vtH_p|}
\vtH_p,
\label{eq:alignment-anisotropic}
\end{equation}
where $\epsilon_p=\pm1$ for alignment and anti-alignment, respectively,
\begin{eqnarray}
n_\nu^\tot\equiv\sum_p n_p,\\
\mu_{\bhp\cdot\bhp^\prime}(n_\nu^\tot) \equiv
\sqrt{2}\GF(1-\bhp\cdot\bhp^\prime) n_\nu^\tot,
\end{eqnarray}
and
\begin{equation}
\langle\vtP_{\bhp}\rangle\equiv
\frac{1}{n_\nu^\tot}\sum_{p^0} n_p \vtP_p
\end{equation}
is the polarization vector averaged across the neutrino/antineutrino
energy spectrum and depends on $\bhp$.
Averaging \eref{eq:alignment-anisotropic} over $p^0$ we obtain
\begin{equation}
\langle\vtP_{\bhp}\rangle
=\frac{1}{n_\nu^\tot}\sum_{p^0}
\frac{\epsilon_p\, n_p |\vtP_p|}{|\vtH_p|}\vtH_p.
\label{eq:avg-vP}
\end{equation}
According to \eref{eq:vtH-anisotropic}, $\vtH_p$ depends on
$\langle\vtP_{\bhp}\rangle$, not on each individual $\vtP_p$. Therefore, 
\eref{eq:L0-cons-anisotropic}, \eref{eq:Lv-cons-anisotropic} and
\eref{eq:avg-vP} are a closed set of $(3\times N_\mathrm{ang}+4)$ coupled
nonlinear integral equations from which we can solve for
$\{\Omega,\bfK,\langle\vtP_{\bhp}\rangle|\forall \bhp\}$ 
given a specified set of parameters
$\{\lambda,|\vP_p|=|\vtP_p|,n_p,\epsilon_p|\forall p\}$. Here 
\begin{equation}
N_\mathrm{ang}\equiv\sum_{\bhp} 1
\end{equation}
is the number of neutrino (angular) trajectories. 

As in the stationary, homogeneous and isotropic case, we are able to
sum out the energy modes in obtaining the precession solution and,
therefore, reduce the number of equations in the closed set by 
a factor of $\sim N_\mathrm{en}$. Nevertheless,
\eref{eq:avg-vP} can become very difficult to solve if $N_\mathrm{ang}$
is more than a few. We also note that, in a stationary, homogeneous
and anisotropic environment, 
for a precession solution $\langle\vP_{\bhp}(t,\bfx)\rangle$ and
$\vP_p(t,\bfx)$ 
do not necessarily lie
in the same plane as they would in an isotropic environment. 

\subsection{Slowly-varying, anisotropic environments%
\label{sec:adiabatic-anisotropic}}

Finally, we consider a neutrino gas in an environment where both
$\lambda(t,\bfx)$ and $n_p(t,\bfx)$ vary slowly with time and/or
space. Like the slowly-varying, homogeneous and isotropic case, we
expect the collective precession mode for neutrino oscillations to be
of the form
\begin{equation}
\fl
\sfP_p(t,\bfx)=\exp\left[-\rmi\phi(t,\bfx)\frac{\Lambda_3}{2}\right]
\stP_p(t,\bfx)
\exp\left[\rmi\phi(t,\bfx)\frac{\Lambda_3}{2}\right].
\label{eq:prec-sol-var}
\end{equation}
In \eref{eq:prec-sol-var}
$\stP_p(t,\bfx)=\stP_p(\lambda(t,\bfx),n_{p^\prime}(t,\bfx))$ has a
weak dependence on 
time and space which arises from the matter density and neutrino number
densities. The set of quantities
$\{\phi(t,\bfx),\stP_p(t,\bfx))|\forall p\}$ 
is a solution to the following equations:
\begin{eqnarray}
\fl
\left[\sfH^\ext_{p^0}
+\sfH^{\nu\nu}_{\bhp}(n_{p^\prime}(t,\bfx),\stP_{p^\prime}(t,\bfx))
-(\partial_t+\bhp\cdot\bnab)\phi(t,\bfx)\frac{\Lambda_3}{2},\,
\stP_p(t,\bfx)\right]=0,
\label{eq:prec-ansatz-var}\\
\partial_t L^0(t,\bfx)+\bnab\cdot\bi{L}(t,\bfx) = 0,
\label{eq:L-cons-var}
\end{eqnarray}
where
\begin{eqnarray}
L^0(t,\bfx)=\sum_p n_p(t,\bfx)\Tr\left[\stP_p(t,\bfx)\,
\Lambda_3\right],\\
\bi{L}(t,\bfx)=\sum_p \bhp\, n_p(t,\bfx)\Tr\left[\stP_p(t,\bfx)\,
\Lambda_3\right].
\label{eq:Lv}
\end{eqnarray}
As discussed in \sref{sec:stationary-anisotropic} equation set
\eref{eq:prec-ansatz-var} can be reduced by a factor of
$N_\mathrm{en}$ by summing it across the neutrino/antineutrino
energy spectrum.

For the collective precession mode we expect that all polarization
vectors $\vP_p(t,\bfx)$ precess collectively about $\ve_3$. This
collective precession is fully described by $\phi(t,\bfx)$ in
\eref{eq:prec-sol-var}, and $\vtP_p(t,\bfx)$ must not rotate about
$\ve_3$. This additional constraint makes \eref{eq:prec-ansatz-var} and
\eref{eq:L-cons-var} generally unsolvable unless all $\vtP_p(t,\bfx)$
lie in the same plane, just as in the homogeneous, isotropic
case. Because \eref{eq:prec-ansatz-var} determines $\phi(t,\bfx)$
up to an arbitrary constant $\phi_0$, we choose an appropriate value
of $\phi_0$ so that 
\begin{equation}
\Tr[\stP_p(t,\bfx)\,\Lambda_2]\equiv 0
\end{equation}
for any neutrino mode $p$ at any $(t,\bfx)$.

It will prove to be helpful to explore static systems in more detail.
In static systems
all physical quantities including polarization vectors are independent
of time $t$. In such a system the collective precession mode
\eref{eq:prec-sol-var} describes a wavy distribution of neutrino
polarization $\sfP_p(\bfx)$. If $\bfK(\bfx)=\bnab\phi(\bfx)$ is constant, then
$\vP_p(\bfx)$ rotates about $\ve_3$ clockwise for a complete cycle as the
neutrino travels along its world line for a distance of
$2\pi/|\bhp\cdot\bfK|$. (In the normal mass hierarchy case, the polarization
vector $\vP_p(t,\bfx)$ for a neutrino with $p^0>0$ rotates about
$\ve_3$ counterclockwise along its world line.)
To gain some insight into the vector $\bfK(\bfx)$ we sum
\eref{eq:prec-ansatz-var} over all neutrino modes and obtain
\begin{equation}
\bfK(\bfx)\cdot\sum_p \bhp\,n_p(\bfx)\tP_{p,1}(\bfx)
=\sum_p [\lambda(\bfx)-\omega_{p^0}] n_p(t,\bfx) \tP_{p,1}(\bfx).
\label{eq:K-dir}
\end{equation}
\Eref{eq:K-dir} shows that $\bfK(\bfx)$ describes an average
oscillation behaviour for neutrinos propagating along some average
direction characteristic of the neutrino 
(lepton) flux. The direction of $\bfK(\bfx)$ can be  determined easily
if the system is fully symmetric about an axis, e.g., the
$\hat{\bi{z}}$ axis, at
$\bfx$. In this case $\bfK(\bfx)\cdot\bhp$, the angular precession frequency
of any neutrino propagating along the direction $\bhp$, must be the
same as that for neutrinos propagating along a different direction
$\bhp^\prime$ as long as
$\bhp\cdot\hat{\bi{z}}=\bhp^\prime\cdot\hat{\bi{z}}$. This means that
$\bfK(\bfx)$ must be parallel to $\hat{\bi{z}}$.

We note that the collective precession mode discussed here
is different from the self-maintained coherence
of neutrino oscillations in the non-spherical geometry which is discussed
in \cite{Dasgupta:2008cu}. What is proposed in \cite{Dasgupta:2008cu}
is based on the assumption that all neutrino flavour polarization vectors
$\vP_p(\bfx)$ are perfectly aligned or anti-aligned with each
other. This assumption forms the basis of the single-angle
approximation in the non-spherical geometry which allows the
computation of neutrino flavour evolution 
along the ``streamlines''. Here streamlines are aligned along the
direction of the neutrino number flux
\begin{equation}
\bi{F}(\bfx)\equiv \sum_p \bhp\,n_p(\bfx).
\end{equation}
In the collective precession mode, polarization vectors are not
required to be aligned (and, in fact, cannot be perfectly aligned)
with each other. 
The vector $\bfK(\bfx)$ is generally not parallel to the neutrino
streamlines, either.

Because it takes little time for neutrinos to traverse 
the region of high neutrino fluxes in supernovae,
all current numerical calculations for supernova neutrino oscillations
are carried out as if all physical conditions such as neutrino fluxes and
the matter profile are static. Computations with various physical
conditions that correspond to supernova evolution over time can
be pieced together to give a dynamic picture of neutrino flavour
transformation in supernovae
(e.g., \cite{Schirato:2002tg,Kneller:2007kg}). 
Generally, the static collective mode parameters computed
with physical inputs at various epochs will not be the
same. This will give a time dependence to $\phi(t,\bfx)$
which actually could, at least in principle, be derived from
\eref{eq:prec-ansatz-var} and \eref{eq:L-cons-var}.
The dynamic collective precession mode \eref{eq:prec-sol-var}
describes a wave-like distribution of 
neutrino polarization whose ``phase'' $\phi(t,\bfx)$ travels with velocity
\begin{equation}
\bi{V}(t,\bfx)\equiv \Omega(t,\bfx)
\frac{\bfK(t,\bfx)}{|\bfK(t,\bfx)|^2},
\end{equation}
where $\Omega(t,\bfx)=-\dot\phi(t,\bfx)$.
It is clear that the static assumption is valid if
\begin{equation}
\fl
\frac{|\Omega|}{|\bfK|}
\simeq (1.4\times10^{-5})
\left(\frac{\tau_\mathrm{dyn}}{1\,\mathrm{s}}\right)^{-1}
\left(\frac{\dm}{3\times10^{-3}\,\mathrm{eV}^2}\right)^{-1}
\left(\frac{E_\star}{5\,\mathrm{MeV}}\right)
\ll 1.
\label{eq:static-cond}
\end{equation}
In \eref{eq:static-cond} we have taken $|\Omega|\simeq2\pi/\tau_\mathrm{dyn}$ and
$|\bfK|= \dm/(2E_\star)$, where $\tau_\mathrm{dyn}$ is the
typical time scale for the variation of the relevant physical
conditions in supernovae. 

\subsection{Criteria for collection neutrino oscillations%
\label{sec:criteria}}

In \sref{sec:adiabatic-anisotropic} we have assumed that at any
space-time point the collective precession mode for neutrino
oscillations is the same as that in a stationary, homogeneous
environment. Therefore, the collective precession mode in a
time-varying, inhomogeneous environment can be established only if the
variation of the physical conditions are ``gentle'', so
that the collective precession mode derived from the stationary,
homogeneous approximation in the neighborhoods of different space-time
points can be connected smoothly. To be more specific, we require that
\begin{equation}
\frac{|(\partial_t+\bhp\cdot\bnab)\theta_p(t,\bfx)|}%
{|\vtH_p(t,\bfx)|}
\ll 1,
\label{eq:adiabatic-cond}
\end{equation}
where
\begin{equation}
\theta_p(t,\bfx)\equiv\arccos\left(
\frac{\vtH_p(t,\bfx)\cdot\ve_3}{|\vtH_p(t,\bfx)|}\right).
\end{equation}
\Eref{eq:adiabatic-cond} is similar to the adiabatic condition used
for the collective precession mode in homogeneous, isotropic neutrino
gases \cite{Raffelt:2007cb}.
Unlike the adiabatic condition used in the MSW mechanism,
\eref{eq:adiabatic-cond} (or the adiabatic condition in
\cite{Raffelt:2007cb}) is applicable only after the adiabatic precession
solution has been determined. This is because neutrinos themselves
contribute to the total flavour evolution Hamiltonian and,
therefore, help set the adiabatic condition.

A more practical criterion, which leads to
a necessary condition for significant 
collective neutrino oscillations to occur, can be
obtained by comparing the magnitudes of
\begin{equation}
\fl
\stH_p^\ext(\bfx)
\equiv\sfH_{p^0}^\ext(\bfx)-\bhp\cdot\bfK(\bfx)\frac{\Lambda_3}{2}
=[\lambda(\bfx)-\omega_{p^0}-\bhp\cdot\bfK(\bfx)]\frac{\Lambda_3}{2}
\label{eq:tHext}
\end{equation} 
and 
\begin{equation}
\fl
  \stH_{\bhp}^{\nu\nu}(\bfx)\equiv
  \sfH_{\bhp}^{\nu\nu}(n_{p^\prime}(\bfx),\stP_{p^\prime}(\bfx))
=\sqrt{2}\GF\sum_{p^\prime}(1-\bhp\cdot\bhp^\prime)
n_{p^\prime}(\bfx)\stP_{p^\prime}(\bfx).
\label{eq:tHnunu}
\end{equation}
Here we have made the static assumption. If 
\begin{equation}
|\stH_p^\ext(\bfx)| \gg |\stH_{\bhp}^{\nu\nu}(\bfx)|,
\label{eq:suppress-cond}
\end{equation}
then \eref{eq:prec-ansatz-var} requires that $\vtP_p(\bfx)$ is either
aligned or antialigned with $\ve_3$. In other words, 
no significant flavour oscillations can occur 
in this case, even if neutrinos are in the
collective precession mode.
We can get a crude estimate for \eref{eq:tHext} and \eref{eq:tHnunu}
if all the $\vtP_p(\bfx)$ are taken to be aligned or
antialigned with one another and we take [see \eref{eq:K-dir}]
\begin{equation}
\fl
\bfK(\bfx)\simeq\left(\sum_p
    [\lambda(\bfx)-\omega_{p^0}] n_p(t,\bfx) \epsilon_p |\vtP_{p}(\bfx)|\right)
\frac{\sum_p \bhp\,n_p(\bfx)\epsilon_p|\vtP_{p}(\bfx)|}%
{\left|\sum_p \bhp\,n_p(\bfx)\epsilon_p|\vtP_{p}(\bfx)|\right|^2}
\label{eq:K-estimate}
\end{equation} 
in computing $\stH_p^\ext(\bfx)$,
where $\epsilon_p=+1$ ($-1$) if
$\vtP_p(\bfx)\cdot\ve_3>0$ ($\vtP_p(\bfx)\cdot\ve_3<0$) near the
neutrino source.
If \eref{eq:suppress-cond} is satisfied for
most neutrinos in some region, then no significant collective neutrino
oscillations are expected to occur in that region.

In  \eref{eq:suppress-cond}, $|\stH_p^\ext(\bfx)|$ measures how
big the difference is between the collective precession frequency
along the world 
line of a neutrino in mode $p$ 
and the intrinsic precession frequency of the neutrino
 when there is no neutrino self-interaction.
On the other hand,
$|\stH_{\bhp}^{\nu\nu}(\bfx)|$ measures the strength of the neutrino
self-interaction which makes collective neutrino oscillations
possible. Therefore, \eref{eq:suppress-cond} can be intuitively
understood as the condition under which neutrino self-interaction is not
strong enough to 
entrain a neutrino mode $p$ whose intrinsic precession frequency is too
different from the collective one.

The condition in
\eref{eq:suppress-cond} implies that a
large matter density can suppress collective neutrino
oscillations in anisotropic environments. 
This is in contrast to the expected behaviour in the homogeneous,
isotropic case, or a supernova model where the single-angle
approximation is employed. 
Matter density-driven suppression of collective oscillations can be
understood as follows.
Using \eref{eq:K-estimate} and assuming $\vtP_p(\bfx)$ to be
either aligned or antialigned with $\ve_3$ (i.e., neutrinos are in
pure flavour states) we obtain
\begin{equation}
\bfK(\bfx)\simeq \sqrt{2}\GF n_e(\bfx) L^0(\bfx) 
\frac{\bi{L}(\bfx)}{|\bi{L}(\bfx)|^2},
\label{eq:K-estimate2}
\end{equation}
where we have ignored vacuum oscillation frequencies.
From \eref{eq:tHnunu} we have
\begin{equation}
\stH_{\bhp}^{\nu\nu}\simeq\sqrt{2}\GF
[L^0(\bfx)-\bhp\cdot\bi{L}(\bfx)]\frac{\Lambda_3}{2}.
\label{eq:tHnunu-estimate}
\end{equation}
Using \eref{eq:tHext}, \eref{eq:K-estimate2}
and \eref{eq:tHnunu-estimate} we can rewrite \eref{eq:suppress-cond} as
\begin{equation}
n_e(\bfx)
\gg
\left|L^0(\bfx)-\bhp\cdot\bi{L}(\bfx)\right|\times
\left|1-\frac{\bhp\cdot\bi{L}(\bfx)}{|\bi{L}(\bfx)|^2}L^0(\bfx)\right|^{-1}.
\label{eq:suppress-cond-ne}
\end{equation}
\Eref{eq:suppress-cond-ne} gives the criterion for a matter density
that is large enough to suppress collective neutrino oscillations in
the anisotropic environment.
Far away from the neutrino source we have $L^0(\bfx)\simeq
|\bi{L}(\bfx)|$ and
\begin{equation}
\fl
n_e(\bfx)\gg
L^0(\bfx)
\simeq [n_{\nu_e}^\tot(\bfx)-n_{\nu_{\mu^\prime}}^\tot(\bfx)
-n_{\bar\nu_e}^\tot(\bfx)+n_{\bar\nu_{\mu^\prime}}^\tot(\bfx)],
\label{eq:suppress-cond2}
\end{equation}
where $n_{\nu_\alpha(\bar\nu_\alpha)}^\tot$
 is the number density of the neutrinos or antineutrinos that are
 initially in the $\alpha$ flavour state at the neutrino source.
The suppression of collective neutrino oscillations by the large matter
density in the supernova environment was first shown in
\cite{EstebanPretel:2008ni}.

\section{Collective neutrino oscillations in supernovae%
\label{sec:sn}}

In \sref{sec:ap-sol} we have demonstrated that the collective
precession mode for neutrino oscillations can arise because of
symmetries in the neutrino flavour evolution equations. 
There is no guarantee that the adiabatic precession solution to
\eref{eq:prec-ansatz-var} and \eref{eq:L-cons-var} exists or that the
corresponding collective oscillation mode is stable. 
Intuitively, however, we do expect such a collective neutrino
oscillation mode to stand out under suitable conditions
while other non-collective oscillation modes decohere kinematically.
In fact, it has been shown \cite{Duan:2006an} that the 
stepwise-spectral-swapping phenomenon can be the
result of collective precession of polarization vectors or
neutrino flavour isospins (NFIS's) when the
single-angle approximation is valid. In this section we present a new
multi-angle simulation which is engineered to have collective neutrino
oscillations occur closer to the neutrino sphere than in previous
simulations. 
Unlike previous calculations, the aspects of this simulation are more
difficult to capture with a single-angle approximation calculation.
Nevertheless, our new simulation provides
compelling evidence for the 
existence of the collective precession mode close to the neutrino
sphere. We will also highlight the
qualitative, multi-angle features in the calculation which can be
explained using the criteria discussed in \sref{sec:criteria}.

\subsection{Collective precession mode%
\label{sec:pr-mode}}

In the new calculation we adopt the same ``neutrino bulb'' model as
in \cite{Duan:2006an} with the radius of the neutrino sphere taken to be
$R=10$ km. We also take similar neutrino energy spectra at the
neutrino sphere, but with $\langle E_{\nu_e}\rangle=10$ MeV, $\langle
E_{\bar\nu_e}\rangle=12$ MeV, 
and $\langle E_{\nu_x}\rangle=\langle E_{\bar\nu_x}\rangle
=13$ MeV, respectively. Here $|\nu_x\rangle$ ($|\bar\nu_x\rangle$)
corresponds to an appropriate linear combination of $|\nu_\mu\rangle$ and
$|\nu_\tau\rangle$ ($|\bar\nu_\mu\rangle$ and $|\bar\nu_\tau\rangle$).  
We take the effective vacuum mixing angle to be
$\thetav=0.1$, and the mass-squared difference to be
$\dm=-3\times10^{-3}\,\mathrm{eV}^2$ (inverted neutrino mass
hierarchy). Note that in the effective $2\times2$ mixing scheme which
we employ, $\thetav\simeq\theta_{13}$ and $|\dm|$ is approximately
the atmospheric neutrino mass-squared difference $\Delta
m^2_\mathrm{atm}$ \cite{Amsler:2008zzb}.
We adopt a simple analytical profile for the electron number density
\cite{Fuller:2005ae}:
\begin{equation}
n_e(r)=(9.2\times10^{30}\,\mathrm{cm}^{-3})
\left(\frac{100}{S}\right)^4\left(\frac{10\,\mathrm{km}}{r}\right)^3,
\end{equation}
where $r$ is the distance to the center of the proto-neutron star, and
the density profile is parametrized by $S$, the entropy per baryon in units
of Boltzmann's constant $k_\mathrm{B}$. In the simulation we discuss
here we take $S=250$.

Because the neutrino bulb model possesses spherical symmetry, the
common azimuthal angle $\phi(r)$ for all NFIS's in the collective
precession mode, if it exists, must depend only on $r$. As a result,  
the NFIS's for various neutrino trajectories 
must precess in phase about $\ve_3$ along $r$.

\begin{figure*}
\includegraphics*[width=\textwidth, keepaspectratio]{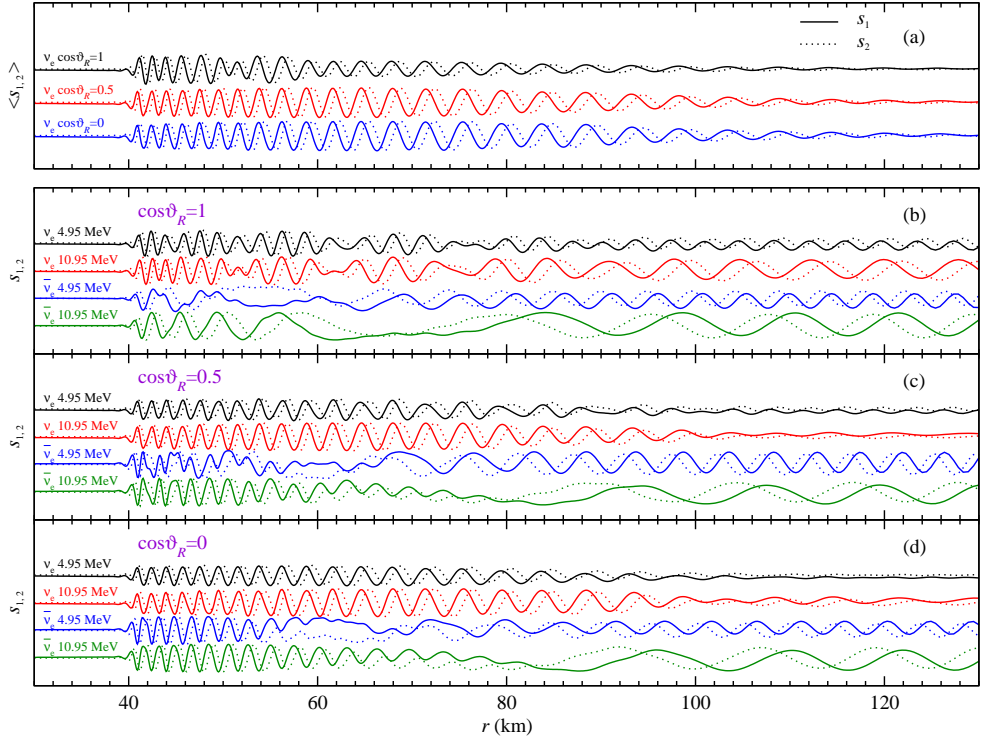}
\caption{\label{fig:sx-r} Oscillations of $s_1(r)$
  (solid lines) and $s_2(r)$ (dotted lines), the 
projection of the NFIS on $\ve_1$ and $\ve_2$, respectively,
as functions of $r$. The curves 
in the top
panel are for neutrinos emitted as $\nu_e$ and propagating along
trajectories with
$\cos\vartheta_R=1$, $0.5$ and $0$, respectively. Here $\langle
s_{1(2)}\rangle$ represents an average over
the initial $\nu_e$ energy spectrum. 
Clearly $\langle\vs(r)\rangle$ for various neutrino trajectories precesses
about $\ve_3$ in phase along $r$.
The curves in the bottom panels
show $s_{1,2}$ for individual energy and angle bins (as labelled). 
The neutrinos or antineutrinos corresponding to these curves start as
pure $\nu_e$ or $\bar\nu_e$, as labelled. 
Note that $s_1(r)$ and $s_2(r)$ are phase-locked at low to modest
radii and again at larger radii.
This suggests that the NFIS's for both
neutrinos and antineutrinos precess with a common frequency at low to modest
radii but precess with energy-dependent vacuum oscillation
frequencies at larger radii. In addition, at large
radii, the NFIS's for
antineutrinos precess in the opposite direction 
with respect to those for neutrinos.} 
\end{figure*}

In \fref{fig:sx-r}(a) we plot $\langle s_{1(2)}(r)\rangle$, the
energy-averaged NFIS component, where 
$s_{1(2)}(r)\equiv\vs_p(r)\cdot\ve_{1(2)}$,
as a function of 
$r$ for neutrinos emitted as pure $\nu_e$ and propagating along a few
representative trajectories.  We here do not distinguish between 
the vacuum mass basis and the
flavour basis because $\thetav\ll1$.
Because of the spherical symmetry, it suffices to
label different neutrino trajectories by
neutrino emission angle $\vartheta_R$, which can be defined to be
\begin{equation}
\cos\vartheta_r(\bhp)\equiv\bhp\cdot\hat{\bfr}
\label{eq:thetar}
\end{equation}
evaluated at $r=R$, with $\hat{\bfr}$ being the radial direction.
(Note that we use $\theta$ and $\vartheta$ to denote angles in 
flavour space and coordinate space, respectively.)
\Fref{fig:sx-r}(a) indeed shows that,
as neutrino oscillations start at $r\simeq40$ km,
 $\langle\vs(r)\rangle$ begin to precess in phase about $\ve_3$ along
$r$. This is a clear signature of the existence of the collective
precession mode in our numerical result.

In figures \ref{fig:sx-r}(b--d) we
plot $s_{1,2}(r)$ for a dozen individual angular and energy bins for both
neutrinos and antineutrinos. 
We observe that $s_1(r)$ and $s_2(r)$ for most neutrinos and
antineutrinos are
phase-locked at low to modest radii and again at larger radii.
The phase-lock at the low to modest radii occurs
because the neutrinos and antineutrinos participate in the collective
precession mode, 
and the phase-lock at larger radii occurs because they experience 
vacuum oscillations. We note that the 
precession of NFIS's in vacuum oscillations is energy
dependent, and can be in the opposite direction to the sense of
precession in the collective
precession mode. This behaviour is especially prominent for the
antineutrino modes in \fref{fig:sx-r}(d).

Two important trends in the
breakdown of the collective precession mode merit further elaboration.
One of these trends is that along each neutrino trajectory antineutrinos 
always drop out of the collective mode earlier than neutrinos, and
antineutrinos with smaller energies drop out earlier than those
with larger energies. This trend can be explained using the criterion in
\eref{eq:suppress-cond}. Using \eref{eq:K-dir} and the spherical
symmetry of the supernova model we obtain
\begin{equation}
K(r)\equiv|\bfK(r)|=\lambda(r) C(r) - \langle\omega(r)\rangle,
\label{eq:K-sn}
\end{equation}
where
\begin{equation}
C(r)\equiv \frac{\sum_p n_p(r) \tP_{p,1}(r)}%
{\sum_p \cos\vartheta_r(\bhp)\, n_p(r) \tP_{p,1}(r)}>1,
\end{equation}
and
\begin{equation}
\langle\omega(r)\rangle\equiv \frac{\sum_p \omega_{p^0} n_p(r) \tP_{p,1}(r)}%
{\sum_p \cos\vartheta_r(\bhp)\, n_p(r) \tP_{p,1}(r)}.
\end{equation}
Combining \eref{eq:tHext} and \eref{eq:K-sn} we have
\begin{eqnarray}
|\stH^\ext_p(r)|&=|\lambda(r)[C(r)\cos\vartheta_r-1]
-[\langle\omega(r)\rangle\cos\vartheta_r-\omega_{p^0}]|,
\\
&\simeq |\omega_{p^0}-\langle\omega(r)\rangle|
\qquad\textrm{if } r\gg R.
\label{eq:tHext-sn}
\end{eqnarray}
Note that in the inverted neutrino mass hierarchy case 
considered here,
we have $\omega_{p^0}=\dm/(2p^0)<0$ for neutrinos ($p^0>0$) and
$\omega_{p^0}>0$ for antineutrinos ($p^0<0$).
Because there are more neutrinos than antineutrinos, we expect
$\langle\omega(r)\rangle<0$ in this calculation. 
From \eref{eq:tHext-sn} one sees that $|\stH^\ext_p(r)|$ is generally
larger for antineutrinos, and the smaller the energy of the
antineutrino is, the larger $|\stH^\ext_p(r)|$ is. Therefore, along the
same neutrino trajectory (and with the same value of
$|\stH^{\nu\nu}_{\vartheta_R}(r)|$), \eref{eq:suppress-cond} is
satisfied at lower radii
for the antineutrinos with smaller energies.
These antineutrinos must drop out of the
collective precession mode earlier than other neutrinos do.

The second trend is that neutrinos and antineutrinos propagating along
the radial trajectory ($\cos\vartheta_R=1$) 
drop out of the collective precession mode earlier than those with the same
energies but propagating along the
tangential trajectory ($\cos\vartheta_R=0$). This can also
be explained using criterion \eref{eq:suppress-cond}. 
As a crude estimate of the strength of the neutrino self-interaction
we can assume that NFIS's are aligned or
antialigned with each other and, therefore,
\begin{eqnarray}
|\stH^{\nu\nu}_{\vartheta_R}(r)|
&\propto
1-\sqrt{1-\left(\frac{R}{r}\right)^2}-
\frac{1}{2}\left(\frac{R}{r}\right)^2\cos\vartheta_r,\\
&\simeq\frac{1}{2}\left(\frac{R}{r}\right)^2(1-\cos\vartheta_r)
+\frac{1}{8}\left(\frac{R}{r}\right)^4
\quad\textrm{if } \left(\frac{R}{r}\right)^2\ll1.
\end{eqnarray}
Clearly $|\stH^{\nu\nu}_{\vartheta_R}(r)|$ is the weakest along the
radial trajectory ($\cos\vartheta_r=\cos\vartheta_R=1$) for which
\eref{eq:suppress-cond} is satisfied first for a given neutrino
energy. 

\subsection{Stepwise spectral swapping%
\label{sec:swap}}

\begin{figure*}
$\begin{array}{@{}c@{\hspace{0.02 \textwidth}}c@{}}
\includegraphics*[scale=0.38, keepaspectratio]{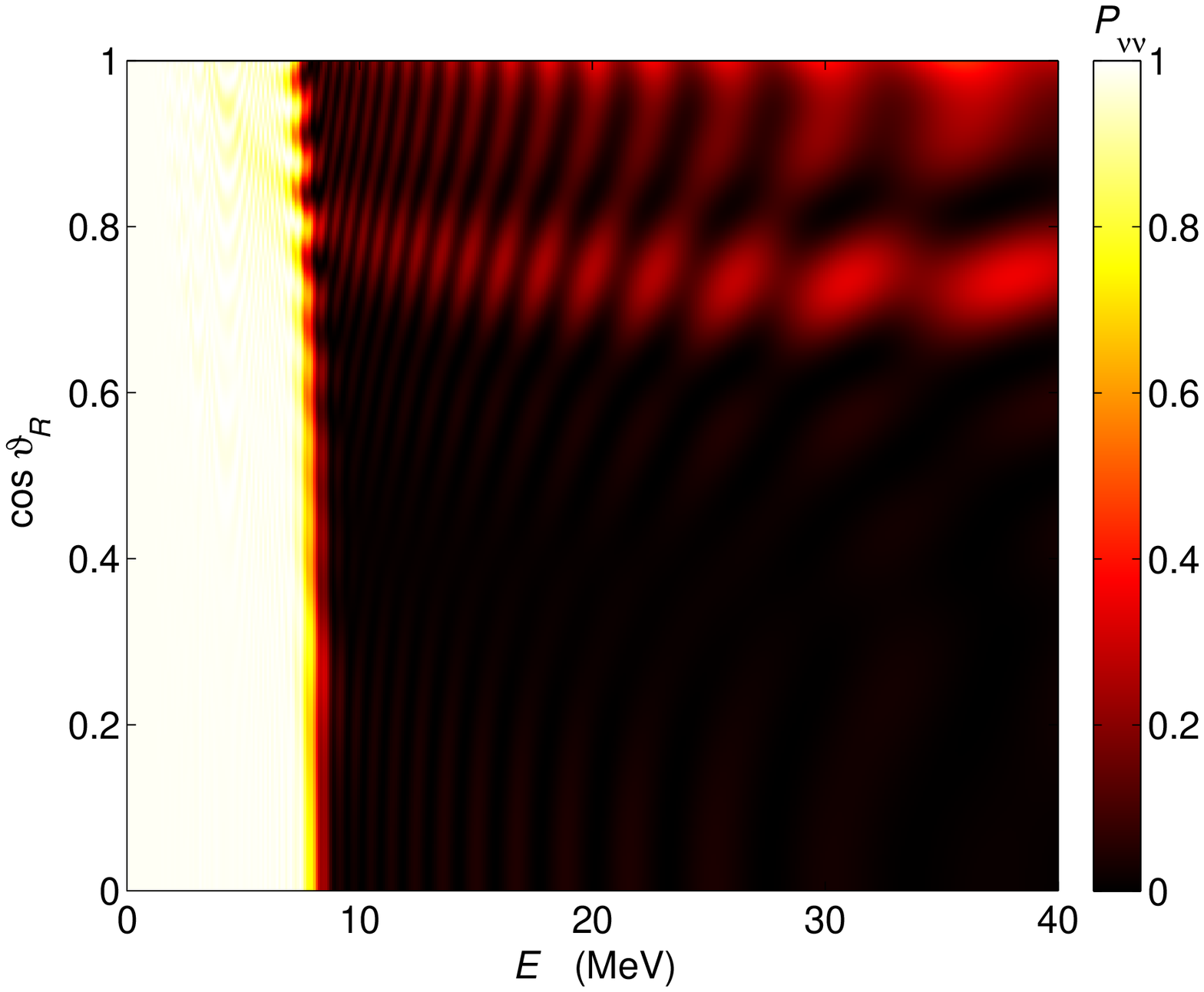} &
\includegraphics*[scale=0.38, keepaspectratio]{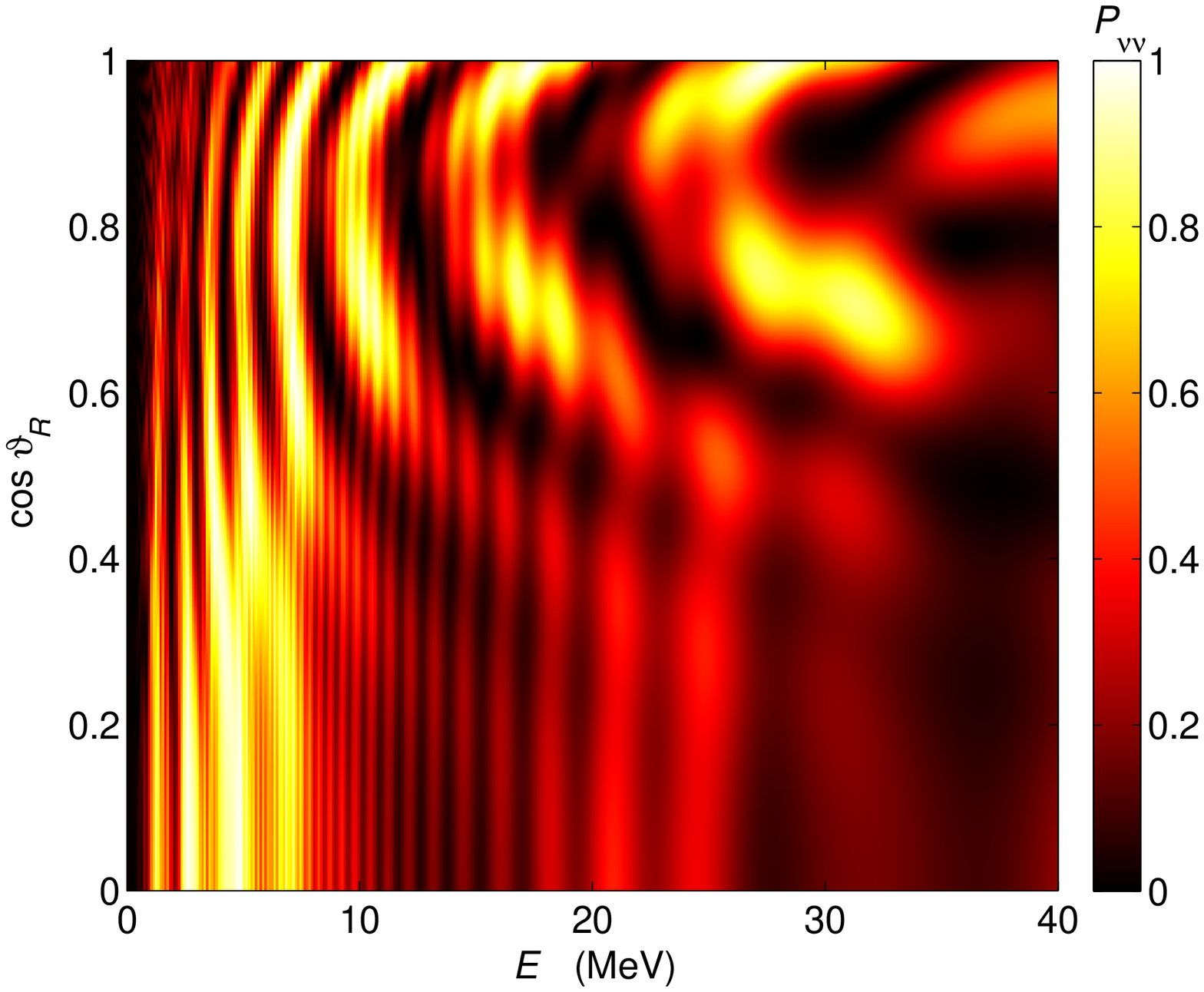}
\end{array}$
\caption{\label{fig:P-E}
The neutrino survival probabilities $P_{\nu\nu}(E,\vartheta_R)$ as functions of
neutrino energy $E$ and emission angle $\vartheta_R$ at $r=200$ km. The
left panel is for neutrinos and the right panel is for antineutrinos. 
Shading code for $P_{\nu\nu}(E,\vartheta_R)$ is given at right.}
\end{figure*}

In \fref{fig:P-E} we plot survival probabilities
$P_{\nu\nu}(E,\vartheta_R)$ at $r=200$ km as functions of both
neutrino energy $E$ and emission angle $\vartheta_R$
for both neutrinos and antineutrinos. 
As in previous calculations, \fref{fig:P-E} shows that $\nu_e$ and
$\nu_x$ swap their energy spectra at energies above $\sim 8$ MeV, a
phenomenon termed as ``stepwise spectral swapping'' or 
``spectral split''. 

For the single-angle approximated supernova model
or for homogeneous, isotropic neutrino gases the 
stepwise-spectral-swapping phenomenon can be explained as the result
of the collective precession 
mode \cite{Duan:2006an,Duan:2007mv,Raffelt:2007cb}. 
If all neutrinos in a homogeneous, isotropic neutrino gas remain in the
collective precession mode to the very end, then the polarization
vector becomes
\begin{equation}
\vP_{p^0} = \epsilon_{p^0}\,
\mathrm{sgn}(\omega_\mathrm{pr}^0-\omega_{p^0})\,
\ve_3,
\label{eq:P-alignment-0}
\end{equation}
where $\epsilon=+1$ ($-1$) if $\vP_{p^0}(t)$ is aligned (antialigned)
with $\vtH_{p^0}(t)$, and $\omega_\mathrm{pr}^0$ is the common
angular precession velocity of $\vP_{p^0}(t)$ about $\ve_3$ when
$n_\nu^\tot\rightarrow 0$. As a result, the neutrino survival
probability becomes
\begin{equation}
P_{\nu\nu}(E)\simeq \left\{\begin{array}{ll}
1 & \textrm{ for neutrinos with }E<\Es,\\
0 & \textrm{ for neutrinos with }E>\Es,\\
0 & \textrm{ for antineutrinos},
\end{array}\right.
\label{eq:Pnunu}
\end{equation}
where
\begin{equation}
\Es=\left|\frac{\dm}{2\omega_\mathrm{pr}^0}\right|.
\label{eq:Es}
\end{equation}

The stepwise-spectral-swapping phenomenon in an anisotropic
environment, such as shown in \fref{fig:P-E}, can be understood through
the collective precession mode discussed in \sref{sec:adiabatic-anisotropic}
in a way similar to above argument based on 
\eref{eq:P-alignment-0}--\eref{eq:Es}, except for the 
replacement $\omega_\mathrm{pr}^0\rightarrow-K|_{r\rightarrow\infty}$. In reality,
however, neutrinos and antineutrinos are not indefinitely entrained in
the collective precession mode. 
The breakdown in collective behaviour results in some interesting
features as shown in \fref{fig:P-E}. For example,
as explained in \sref{sec:pr-mode}, 
antineutrinos, especially those with low energies and/or propagating
along the radial trajectory, do not
participate as well in the collective mode as other neutrinos do. Therefore,
the simplistic prescription in equation \eref{eq:Pnunu} 
works better for neutrinos than
for antineutrinos. In fact, according to \fref{fig:P-E}, this
prescription breaks down for antineutrinos 
 with energies $\sim5$ MeV and/or for those propagating along the
radial trajectory.  In
addition, antineutrinos with energies $E\lesssim0.8$
MeV do not participate in collective oscillations at all and are
almost fully converted through the conventional matter-driven MSW
mechanism. Antineutrinos with 
energies $E\sim5$ MeV do participate in collective
oscillations. However, they encounter MSW resonances after
they leave the collective precession mode and, as a result, are
(partially) converted back to their original flavours.

Another important detail in \fref{fig:P-E} is that the critical
energy $\Es$ is actually 
different for neutrinos propagating along different 
trajectories.
In contrast, $K(r)$ is the same along these trajectories.
Two main factors contribute to
the variation of $\Es$ from trajectory to trajectory.
The first
factor is that the
collective precession frequency $K(r)\cos\vartheta_r$ along a neutrino
world line depends on the neutrino propagation direction.
It is largest along the radial trajectory and the smallest
along the tangential trajectory.
The second factor is that
neutrinos propagating along different trajectories drop out of the
collective precession mode at different radii. 
Neutrinos propagating along more radially-directed trajectories stop
participating in collective oscillations at smaller distances from the
proto-neutron star, but
neutrinos propagating along more tangential trajectories stay in collective
oscillations out to larger distances.
Our numerical calculations show that
the collective precession frequency always decreases as 
neutrino fluxes decrease. Therefore, both contributions lead to a smaller
$\Es$ for the radial trajectory and a larger $\Es$ for the tangential
trajectory. Indeed, the left panel of \fref{fig:P-E} shows that
$\Es$ for the radial and tangential trajectories differ by
$\sim0.7$ MeV.

\section{Conclusions%
\label{sec:conclusions}}

We have demonstrated the existence of the SU($N_\mathrm{f}$) rotation
symmetry in the neutrino flavour evolution equations. 
This symmetry can facilitate the 
establishment of the collective precession mode for
neutrino flavour oscillations in various environments.
The stepwise-spectral-swapping phenomenon can
develop from such a collective neutrino oscillation mode in, e.g.,
supernovae. We have also given criteria for significant neutrino
flavour oscillations to occur if neutrinos are entrained in the
collective precession mode. These criteria can be used to
understand the suppression of collective neutrino oscillations in 
anisotropic environments in the presence of a large matter density
\cite{EstebanPretel:2008ni}.
These criteria also illuminate the process of
 the breakdown of collective oscillations when neutrino densities
are low. The results obtained in both our new simulation and 
previous multi-angle calculations for supernova neutrino oscillations
can be understood in terms of the collective precession mode and the
criteria we have provided.

There remains much to be learned about the collective precession mode
for neutrino oscillations. This collective mode can not exist when
there is no solution to \eref{eq:prec-ansatz-var} and
\eref{eq:L-cons-var}.  
A related interesting observation is that collective oscillations, if
any, for a symmetric system with equal numbers of neutrinos and
antineutrinos are quickly disrupted 
in the presence of even an infinitesimal anisotropy and flavour
equipartition is obtained as a result \cite{Raffelt:2007yz}.  
In the flavour pendulum analogy
\cite{Hannestad:2006nj}, the asymmetry of neutrinos and antineutrinos
constitutes the internal spin of the flavour pendulum. 
The existence of this internal spin causes the flavour pendulum to undergo
the precession which represents the collective precession mode for
neutrino oscillations. There exists no collective
precession mode in a symmetric neutrino system --- this explains the
finding in \cite{Raffelt:2007yz}. Meanwhile, it has been shown in
\cite{EstebanPretel:2007ec} that a typical asymmetry in the neutrino and
antineutrino fluxes in the supernova environment 
 will suppress  multi-angle decoherence
and, therefore, make the collective precession mode for neutrino
oscillations possible.

\ack
This work was supported in part by 
DOE grants DE-FG02-00ER41132 at the INT,
DE-FG02-87ER40328 at UMN,
NSF grant PHY-06-53626 at UCSD,
and an IGPP/LANL mini-grant.
This research used resources of the National Energy Research
Scientific Computing Center, which is supported by the Office of
Science of the US Department of Energy under Contract
No.\ DE-AC02-05CH11231.
We thank J Carlson, J J Cherry, A Friedland, W Haxton,
D B Kaplan, C Kishimoto,
C Lunardini, A Mezzacappa,
G Raffelt and P Romatschke for insightful discussions.

\section*{References}
\bibliographystyle{iopart-num}
\bibliography{ref}

\end{document}